\documentclass[letterpaper,journal]{IEEEtran}
\usepackage{amsmath,amsfonts}
\usepackage{algorithmic}
\usepackage{algorithm}
\usepackage{array}
\usepackage[caption=false,font=normalsize,labelfont=sf,textfont=sf]{subfig}
\usepackage{textcomp}
\usepackage{stfloats}
\usepackage{booktabs}
\usepackage{url}
\usepackage{verbatim}
\usepackage{graphicx}
\usepackage{multirow}
\usepackage{makecell}
\usepackage{svg}
\usepackage{placeins}
\usepackage[commandnameprefix=always]{changes}
\definechangesauthor[name={Author}, color=blue]{A}
\def\BibTeX{{\rm B\kern-.05em{\sc i\kern-.025em b}\kern-.08em
    T\kern-.1667em\lower.7ex\hbox{E}\kern-.125emX}}

\begin{document}

\title{Bias and Fairness in Self-Supervised Acoustic Representations for Cognitive Impairment Detection}

\author{Kashaf Gulzar,~\IEEEmembership{Member,~IEEE}, 
Korbinian Riedhammer,~\IEEEmembership{Senior Member,~IEEE}, Elmar N\"oth,~\IEEEmembership{Member,~IEEE}, Andreas K. Maier, 
Paula Andrea P\'erez-Toro,~\IEEEmembership{Member,~IEEE}
\thanks{K. Gulzar and K. Riedhammer are with Zentrum für K\"unstliche Intelligenz (KIZ), Technische Hochschule N\"urnberg, N\"urnberg, Germany (email: kashaf.gulzar@th-nuernberg.de; korbinian.riedhammer@th-nuernberg.de).} 
\thanks{E. N\"oth, A. K. Maier and P. A. P\'erez-Toro are with Pattern Recognition Lab, Friedrich-Alexander Universit\"at Erlangen-N\"urnberg, Erlangen, Germany (email: elmar.noeth@fau.de; andreas.maier@fau.de; paula.andrea.perez@fau.de).}
\thanks{P. A. P\'erez-Toro is with GITA Lab, University of Antioquia, Medell\'in, Colombia.}
}

\markboth{}
{K. Gulzar \MakeLowercase{\textit{(et al.)}: 
Bias and Fairness in Self-Supervised Acoustic Representations for Cognitive Impairment Detection}}


\maketitle
\begin{abstract}
Speech-based detection of cognitive impairment (CI) offers a promising non-invasive approach for early diagnosis, yet performance disparities across demographic and clinical subgroups remain underexplored, raising concerns around fairness and generalizability. 
This study presents a systematic bias analysis of acoustic-based CI and depression classification using the DementiaBank Pitt Corpus. 
We compare traditional acoustic features (MFCCs, eGeMAPS) with contextualized speech embeddings from Wav2Vec 2.0 (W2V2), and evaluate classification performance across gender, age, and depression-status subgroups.
For CI detection, higher-layer W2V2 embeddings outperform baseline features (UAR up to 80.6\%), but exhibit performance disparities; specifically, females and younger participants demonstrate lower discriminative power (\(AUC\): 0.769 and 0.746, respectively) and substantial specificity disparities (\(\Delta_{spec}\) up to 18\% and 15\%, respectively), leading to a higher risk of misclassifications than their counterparts.
These disparities reflect representational biases, defined as systematic differences in model performance across demographic or clinical subgroups.
Depression detection within CI subjects yields lower overall performance, with mild improvements from low and mid-level W2V2 layers. 
Cross-task generalization between CI and depression classification is limited, indicating that each task depends on distinct representations.
These findings emphasize the need for fairness-aware model evaluation and subgroup-specific analysis in clinical speech applications, particularly in light of demographic and clinical heterogeneity in real-world applications.
\end{abstract}

\begin{IEEEkeywords}
Alzheimer's disease, cognitive impairment, depression, acoustic embeddings, bias analysis
\end{IEEEkeywords}

% --------------------------------------------------------------------------------------
\section{Introduction}
Cognitive decline (CD) refers to a gradual and often irreversible deterioration of cognitive abilities, including memory, reasoning, attention, and language skills, commonly associated with aging. 
However, various neurological and psychological disorders, notably Alzheimer's disease (AD) and depression, are primary causes of CD, posing significant challenges at both individual and societal levels \cite{roark}.
Within this spectrum, cognitive impairment (CI) is a broader clinical category describing measurable deficits exceeding typical age-related changes. 
Mild cognitive impairment (MCI), a specific early-stage of CI, involves noticeable cognitive difficulties that do not yet severely interfere with daily functioning \cite{grundman}.
MCI holds clinical significance as a transitional phase between normal aging and dementia, particularly AD, with affected individuals facing an elevated risk of progression \cite{roark}.

AD is a progressive neuro-degenerative disorder, primarily marked by memory loss, impaired reasoning, and behavioral changes. 
Early detection, particularly at the MCI stage, is critical for timely intervention and improved patient outcomes. 
However, this remains difficult due to reliance on clinical biomarkers and the subtlety of early symptoms \cite{grundman}. 
Moreover, 30–50\% of AD patients experience depressive symptoms, particularly in early and intermediate CI stages \cite{hochang,alzheimer2018}. 
Overlapping symptoms such as disrupted sleep, apathy, and poor focus make differential diagnosis challenging \cite{Korczyn}. 
Since depression remains treatable and its co-occurrence increases the risk of progression to AD \cite{modrego}, accurate differentiation is clinically crucial.

\subsection{Related Work}
In recent years, there has been growing interest in leveraging machine learning (ML) techniques and acoustic analysis for detecting CI and associated conditions like depression~\cite{konig}.
Existing work can be broadly categorized into three methodological directions: (1) \textit{engineered acoustic feature-based approaches for AD detection}, (2) \textit{acoustic-based classification of depression within CI populations}, and (3) \textit{embedding-based and transfer learning models}.

\IEEEpubidadjcol
A substantial body of work has investigated traditional acoustic features for AD detection. 
The authors in~\cite{zhou} utilized the Alzheimer's dementia recognition through spontaneous speech only (ADReSSo) challenge dataset, applying logistic regression classifiers to a range of acoustic and linguistic features. 
Among various acoustic configurations, the extended geneva minimalistic acoustic parameter set (eGeMAPS) feature set yielded the highest accuracy of 74\%. 
Similarly, Moro-Velázquez \textit{et al.}~\cite{moro} combined mel frequency cepstral coefficients (MFCCs), silence-related features, and neural embeddings within a probabilistic linear discriminant analysis framework, achieving 73\% accuracy for acoustic features alone. 
In another study, Balagopalan and Novikova~\cite{balagopalan} employed MFCCs, jitter, shimmer and zero-crossing rate, alongside multiple ML models such as support vector machines (SVMs), neural networks, and decision trees. 
Their findings demonstrated 66\% accuracy for acoustic features using SVMs.

Parallel research has explored acoustic-based depression classification within CI populations. 
Abdallah-Qasaimeh and Ratté~\cite{abdallahmci} evaluated spectral, prosodic, and MFCCs-derived statistical features from the DementiaBank (DB) Pitt corpus, employing SVM and random forest classifiers. 
Their best performance reached 91\% accuracy with random forest models. 
Expanding on this, Abdallah \textit{et al.}~\cite{abdallahann} organized acoustic features into incremental groups and applied a multi-layer perceptron (MLP), achieving a maximum accuracy of 77\%. 

More recently, embedding-based and transfer learning approaches have emerged. 
Perez-Toro \textit{et al.}~\cite{paula} introduced a transfer learning framework combining prosodic and spectral features from both the DB Pitt corpus and the interactive emotional dyadic motion capture (IEMOCAP) speech datasets. 
Using a ForestNet model, their system achieved unweighted average recall (UAR) scores of 87\% for AD detection and 82\% for depression classification. 
In a complementary study, Braun \textit{et al.}~\cite{braun} conducted a cross-corpus analysis utilizing Wav2Vec 2.0 (W2V2) embeddings and SVMs on two independent German speech datasets. 
Despite achieving moderate performance (UAR=62\%), their work underscored the challenges of cross-dataset generalizability in speech-based dementia detection. 

Alongside performance-driven efforts, concerns regarding fairness and demographic bias in self-supervised learning (SSL) models have increasingly drawn attention.
% Several studies have demonstrated that speech SSL models can encode and propagate biases inherent in training data, resulting in performance disparities across demographic groups.
Studies have shown that models like Wav2Vec 2.0 (W2V2) and Whisper exhibit demographic performance disparities, particularly across gender and age groups~\cite{boito2022study, fuckner2023uncovering, lin2024social}.
Furthermore, a study by Hu et al.~\cite{hu} investigates bias and robustness in SSL models through automatic speech recognition (ASR) performance and downstream binary AD detection analyzing linguistic representations, across elderly and dysarthric speech and multiple languages. 
Despite promising advances, no existing studies have examined the potential influence of demographic (e.g., age, gender) and clinical factors on the performance and fairness of acoustic-based ML models for CI detection. 
This oversight limits the clinical reliability and equity of current systems and underscores the need for bias-aware evaluation in this domain.

\subsection{Contribution of this Study}
In this work, we systematically investigate the impact of demographic and clinical factors on the performance of acoustic-based ML models for CI detection.
Using the DementiaBank Pitt Corpus, we evaluate traditional baseline features (MFCCs, eGeMAPS) alongside Wav2Vec 2.0 embeddings across CI and depression classification tasks. 
Our key contributions are:
\begin{itemize}
    \item We comparatively evaluate baseline features and W2V2 embeddings for CI and comorbid depression classification.
    \item We systematically analyze the influence of demographic and clinical factors on W2V2 performance, uncovering biases impacting fairness and clinical reliability.
    \item We examine representational overlap and cross-task generalization between CI and depression detection.
\end{itemize}

To our knowledge, this is the first comprehensive investigation into demographic and clinical biases in speech-based CI detection. 
Beyond classification performance, our work is motivated by growing concerns about the fairness and reliability of self-supervised models in clinical speech applications. 
Speech-based AI tools are increasingly proposed for early detection of cognitive disorders, yet unequal model performance across demographic or clinical subgroups may risk reinforcing existing health disparities. 
By quantifying these subgroup disparities and highlighting persistent representational biases even after data balancing, we aim to promote fairness-aware evaluation protocols for future clinical deployment of speech-based machine learning systems.

% --------------------------------------------------------------------------------------
\section{Data}
\begin{table*}
\centering
\caption{Demographic and clinical information of subjects for Imbalanced, CI-balanced, and CI-gender balanced datasets. CI: CI patients, NCI: NCI subjects, Age grp: Age group, F: Female, M: Male. MMSE and HAM-D scores are expressed as mean (standard deviation).}
\label{tab:data_distribution}
\begin{tabular}{@{}c|cc|cc|cc@{}}
\toprule
 & \multicolumn{2}{c|}{\textbf{Imbalanced}} & \multicolumn{2}{c|}{\textbf{CI-balanced}} & \multicolumn{2}{c}{\textbf{CI-gender balanced}} \\ \cmidrule(l){2-7} 
 & \textbf{\begin{tabular}[c]{@{}c@{}}CI\\ {[}F/M{]}\end{tabular}} & \textbf{\begin{tabular}[c]{@{}c@{}}NCI\\ {[}F/M{]}\end{tabular}} & \textbf{\begin{tabular}[c]{@{}c@{}}CI\\ {[}F/M{]}\end{tabular}} & \textbf{\begin{tabular}[c]{@{}c@{}}NCI\\ {[}F/M{]}\end{tabular}} & \textbf{\begin{tabular}[c]{@{}c@{}}CI\\ {[}F/M{]}\end{tabular}} & \textbf{\begin{tabular}[c]{@{}c@{}}NCI\\ {[}F/M{]}\end{tabular}} \\ \midrule
Gender & 97/42 & 51/39 & 63/27 & 51/39 & 39/39 & 39/39 \\
Age grp. 1 & 22/13 & 29/25 & 19/10 & 29/25 & 14/12 & 21/25 \\
Age grp. 2 & 75/29 & 22/14 & 44/17 & 22/14 & 25/27 & 18/14 \\
CI-low & 0/0 & 51/39 & 0/0 & 51/39 & 0/0 & 39/39 \\
CI-mild & 75/28 & 0/0 & 56/18 & 0/0 & 34/25 & 0/0 \\
CI-severe & 22/14 & 0/0 & 7/9 & 0/0 & 5/14 & 0/0 \\
Depressed & 33/18 & 3/7 & 24/11 & 3/7 & 13/16 & 3/7 \\
MMSE & \makecell{18.29 (4.12)/\\17.37 (4.29)} & \makecell{29.29 (0.99)/\\28.85 (1.05)} & \makecell{19.40 (3.73)/\\17.52 (4.53)} & \makecell{29.29 (0.99)/\\28.85 (1.05)} & \makecell{19.15 (4.05)/\\17.13 (4.31)} & \makecell{29.44 (0.84)/\\28.85 (1.05)} \\
HAM-D & \makecell{6.69 (3.04)/\\7.30 (4.69)} & \makecell{2.20 (3.11)/\\3.79 (4.26)} & \makecell{6.89 (3.91)/\\7.48 (5.24)} & \makecell{2.20 (3.11)/\\3.79 (4.26)} & \makecell{6.31 (3.71)/\\7.15 (4.63)} & \makecell{2.33 (3.35)/\\3.79 (4.26)} \\ \bottomrule
\end{tabular}%
\end{table*}

The dataset comprises semi-spontaneous speech recordings from 229 participants in a subset of the DementiaBank Pitt corpus~\cite{pittcorpus}, including 139 AD patients with CI and 90 cognitively normal (NCI) controls. 
Participants performed the "Cookie Theft" picture description task~\cite{goodglass1983}.
Recordings were preprocessed by removing interviewer speech using provided timestamps.

Participants were divided into two age groups: 0–65 years (group 1) and 66 years and older (group 2), based on the standard clinical cutoff for early and late onset AD \cite{seath2024clinical}. 
Clinical labels for CI and depression were assigned based on Mini-Mental State Examination (MMSE) and Hamilton Depression Rating Scale (HAM-D) scores, respectively, following established clinical guidelines. 
MMSE scores range from 0 to 30, with scores below 24 indicating CI~\cite{mmse}, while HAM-D scores range from 0 to 22; scores equal to or greater than 8 denote the presence of depression \cite{hamilton1960hamilton, zimmerman2013severity}. Among the AD patients, the non-depressed group (n = 88) had a mean HAM-D score of 4.41, while the depressed group (n = 51) had a mean score of 11.20.

\begin{figure}[H]
    \centering
    \includegraphics[width=\linewidth]{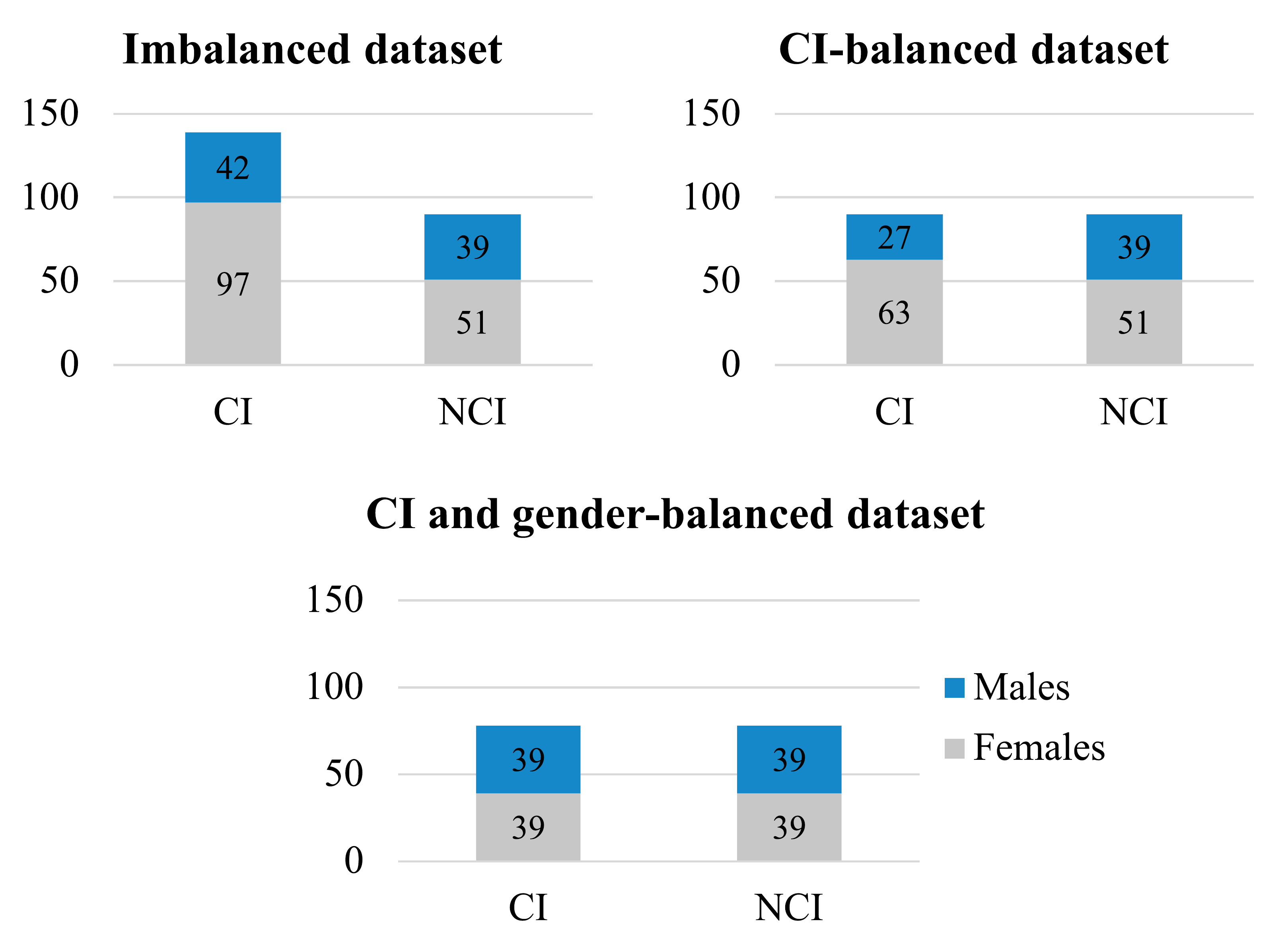}
    \caption{Subject distribution based on CI and gender.}
    \label{fig:data_distribution}
\end{figure}

The dataset exhibited class imbalance. To mitigate this, we balanced the data first by CI status and then by gender during preprocessing (see Figure~\ref{fig:data_distribution}). 
Speakers were randomly sampled from each group, and additional experiments used remaining samples as an alternative test set to validate robustness.

Balancing by age group was not feasible due to the predominance of AD in older adults ($\geq$ 65 years)~\cite{alzheimer2018} and the limited number of younger AD participants, reflecting inherent dataset characteristics (see Table~\ref{tab:data_distribution}).

Table~\ref{tab:data_distribution} details the distribution for the original imbalanced data, CI-balanced data, and CI-gender balanced subsets. 
Participants with MMSE scores above 24 were classified as cognitively normal. 
Due to the small number of depressed individuals in the NCI group, depression-related analyses were limited to the CI participants.

To minimize the impact of confounding factors related to data acquisition, all audio recordings were preprocessed with denoising and normalization procedures. 
These steps were applied consistently across all participants, ensuring that variations in recording conditions do not systematically bias model performance across demographic or clinical subgroups.

% --------------------------------------------------------------------------------------
\section{Methods}
\begin{figure}[H]
    \centering
    \includegraphics[width=\linewidth]{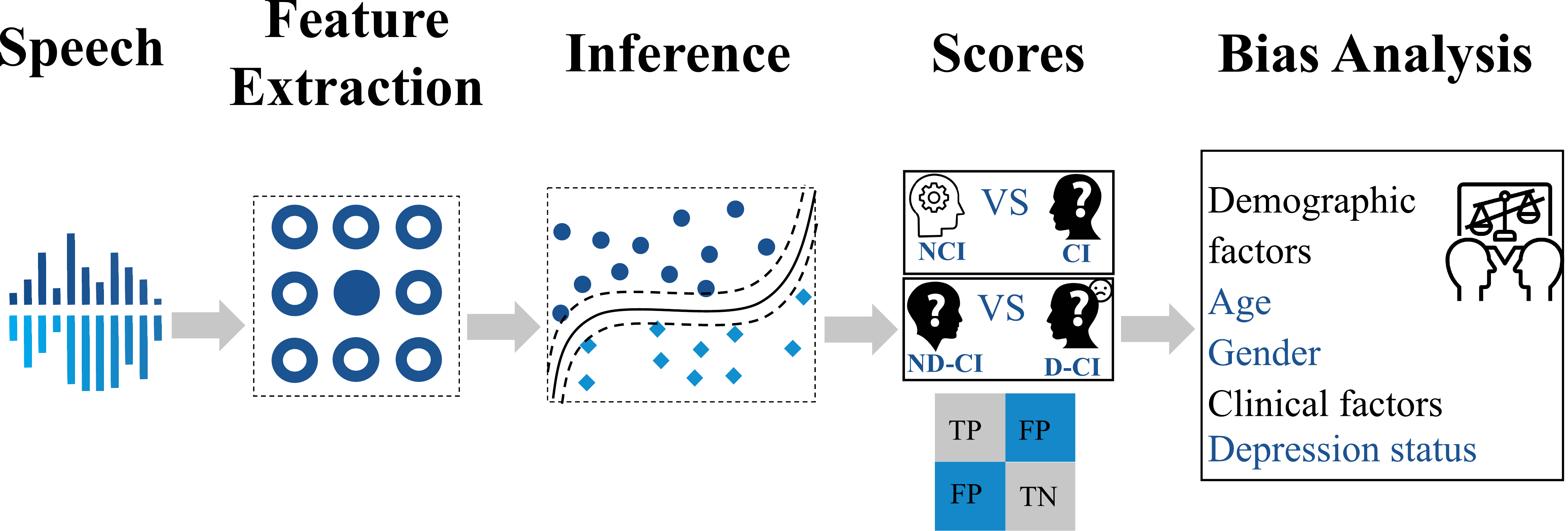}
    \caption{Workflow diagram illustrating the key steps in the data analysis.}
    \label{fig:workflow}
\end{figure}

Figure~\ref{fig:workflow} illustrates the overall workflow. 
Acoustic representations including MFCCs, eGeMAPS, and W2V2 were extracted from speech recordings across imbalanced, CI-balanced and CI-gender balanced datasets.
Two classification tasks were conducted for each feature set: (1) \textit{CI vs. NCI}, and (2) \textit{Depressed CI (D-CI) vs. Non-Depressed CI (ND-CI).}
The second classification task excluded depressed NCI subjects due to their limited number in the dataset (cf. Table~\ref{tab:data_distribution}).
To investigate potential demographic and clinical biases, a bias analysis was carried out for each of the predicted data category, considering factors such as age, gender, and depression status (based on HAM-D).
CI severity was not considered in bias analysis, since low CI severity subjects appeared exclusively in the NCI group, and mild/severe cases only in the CI group.

All data were originally collected by the DementiaBank Pitt Corpus team with informed consent from participants and approval from institutional review boards (IRB). 
This study utilized only the existing de-identified dataset, without any new data collection or direct interaction with participants.

\subsection{Acoustic representations}
We employed three types of acoustic representations varying in complexity: basic spectral features (MFCCs), a further derived parameter set (eGeMAPS), and utterance based neural embeddings (W2V2).
MFCCs and eGeMAPS served as baselines, with W2V2 embeddings used for detailed evaluation and bias analysis. 
All features were independently normalized to ensure consistent scaling.
These representations were selected as they collectively capture both conventional and state-of-the-art approaches to model acoustic patterns relevant to CI and depression classification.

\subsubsection{Mel Frequency Cepstral Coefficients}
MFCCs are widely used speech features inspired by the human auditory system’s non-linear perception of sound. 
They map frequencies onto the Mel scale, which better aligns with human pitch perception~\cite{tiwari}. 
MFCCs are computed by applying a Fast Fourier Transform (FFT) to short-time windows of the speech signal, converting the resulting spectrum to a mel-spectrogram, taking its logarithm, and then applying a Discrete Cosine Transform (DCT). 
These coefficients are known for their robustness to speaker and recording variability~\cite{dave}.

In this study, 40-dimensional MFCCs were extracted using 2048-sample windows with a 512-sample hop at a 16~kHz sampling rate \cite{meghanani2021mfcc}. 
To capture speaker-specific information, the MFCCs were averaged along the time dimension for each coefficient, resulting in a 1-dimensional representation while retaining all 40 MFCC features~\cite{pulido}.
Although dynamic coefficients (\(\Delta \) and \(\Delta \)\(\Delta \)) and additional statistics can be incorporated, internal cross-validation indicated that mean-aggregated MFCCs provided the most effective and stable configuration for our classification tasks, in line with prior work in CI speech analysis \cite{balagopalan}.

\subsubsection{Extended Geneva Minimalistic Acoustic Parameter Set}
eGeMAPS is a standardized set of acoustic features for speech analysis, designed to capture voice characteristics relevant to affective and clinical applications~\cite{eyben1}. 
It includes 88 parameters, extending the minimalistic set of 18 Low-Level Descriptors (LLDs) by adding cepstral and dynamic features~\cite{eyben2}. 

\begin{itemize}
 \item \textbf{Frequency related parameters:} Including pitch, jitter, and the center frequencies and bandwidths of formants 1, 2, and 3.
 \item \textbf{Amplitude related parameters:} Including shimmer, loudness and harmonics-to-noise ratio.
 \item \textbf{Spectral (balance) parameters:} Including alpha ratio, Hammarberg index, spectral slopes (0-500 Hz, 500-1500 Hz), MFCCs 1–4, spectral flux, and harmonic difference (H1–H2, H1-H3).
\end{itemize}

For this study, the latest 'eGeMAPSv02' configuration at the 'functionals' feature level was employed, encompassing features that capture overall speech patterns indicative of cognitive decline~\cite{eyben1}.

\subsubsection{Wav2Vec 2.0}
W2V2 is a self-supervised neural embedding model for speech that combines convolutional front-end layers with transformer blocks to learn contextual speech representations from raw audio~\cite{baevski20}. 
It outputs 512-dimensional embeddings every 20~ms at each transformer layer, which can be aggregated (e.g., via averaging) for downstream tasks.
W2V2 is pre-trained using a contrastive learning framework, predicting masked audio segments based on surrounding context in large, unlabeled speech datasets. 
This minimizes a contrastive loss, encouraging similar representations for neighboring speech segments~\cite{xu}. Its hierarchical architecture enables extraction of both low- and high-level acoustic patterns.

We used the Wav2Vec2-Base-960h model, pre-trained and fine-tuned on 960 hours of 16kHz LibriSpeech data, given its strong performance in speech-based classification~\cite{wav2vec2}. 
Embeddings from both latent and hidden layers were individually assessed for classification.
In this context, latent layers correspond to the outputs of the convolutional feature encoder, which compress raw audio into acoustic representations \cite{baevski20}, while hidden layers refer to the transformer encoder states that capture contextualized, higher-level speech features \cite{pasad2021layer}.

\subsection{Optimization and Classification}
The classification was performed using three classifiers from distinct algorithm families that are frequently used, ensuring comprehensive coverage of bias analysis scenarios: (1) \textit{Radial Basis Function-Support Vector Machine (RBF-SVM)}~\cite{balagopalan}, (2) \textit{RF}~\cite{abdallahmci}, and (3) \textit{MLP}~\cite{abdallahann}.

In case of RBF-SVM, the hyper-parameters were determined via stratified 5-fold cross-validation grid search where $C \in \{ 10^{-2}, 10^{-1}, \ldots, 10^{3} \}$ and $\gamma \in \{ 10^{0}, 10^{-1}, \ldots, 10^{-4} \}$.
Tuning was carried out individually for each experimental setup using the dataset employed in that specific task.
Accuracy was used as the primary optimization metric during model selection.

The RF classifier was designed with an ensemble of 100 decision trees, with each tree grown to its full depth unless a node contained fewer than two samples. 
The number of features considered for splitting at each node was set to square root of the total number of input features.
To ensure reproducibility, a fixed random seed (42) was applied in all classification experiments.
The ensemble nature of RF typically mitigates over-fitting, making it less sensitive to specific hyper-parameter adjustments compared to models such as SVMs.
Consequently, the RF configuration was chosen based on common practice in AD classification studies~\cite{RF_config}, as the primary focus of this study was not to optimize the RF hyper-parameters but rather to examine bias in acoustic representations.

The MLP comprised three hidden layers with 150, 100, and 50 neurons, respectively, arranged to reduce complexity and improve generalization.
The Rectified Linear Unit (ReLU) activation function was applied to the hidden layers, and optimization was performed using the Adam optimizer with a learning rate of 0.001.
The number of epochs was set to 1,000 to achieve optimal classification performance, and a random seed of 42 was applied to maintain reproducibility.
The MLP model configuration was selected based on the architecture used in~\cite{MLP_architecture} for AD classification, as the primary focus of this study was not to optimize the MLP architecture, but rather to compare the performance of three different algorithm families.

For all the classifiers, the data was split into 70\% training and a 30\% test sets. 
All splits were stratified based on the classification labels: CI status for the CI vs. NCI task and depression status for the D-CI vs. ND-CI task.
Final evaluation metrics were obtained by averaging results across five fixed random data splits with seeds 0, 50, 100, 150, and 200.

To support reproducibility, the source code, and lists of audio filenames corresponding to each experimental condition (imbalanced, CI-balanced, and CI-gender balanced) are publicly available on our GitHub repository: \url{https://github.com/th-nuernberg/taslp-bias-in-dementia}.
Due to restrictions on data sharing, the raw audio recordings and transcripts from the DB Pitt Corpus cannot be distributed by us. 
Researchers interested in the dataset can request access directly from DementiaBank at\footnote{\url{https://dementia.talkbank.org}}.

\subsection{Fairness Analysis}
To assess potential model-related biases, we conducted a subgroup-wise performance evaluation across key demographic and clinical factors, including age group, gender, and depression status. 
For each subgroup, we computed subgroup-specific sensitivity and subgroup-specific specificity to allow fair comparison independent of global dataset imbalance.
\[
\parbox{0.35\columnwidth}{\centering
Subgroup-Specific\\Sensitivity}
=
\frac{\text{Correct CI predictions in subgroup}}{\text{CI samples in that subgroup}}
\]
\[
\parbox{0.35\columnwidth}{\centering
Subgroup-Specific\\Specificity}
=
\frac{\text{Correct NCI predictions in subgroup}}{\text{NCI samples in that subgroup}}
\]

Intra-group performance imbalance (\(\delta\)) for a given subgroup \textit{g} was quantified as the difference between subgroup-specific sensitivity and subgroup-specific specificity:
\[
\parbox{0.10\columnwidth}{\centering
\(\delta_g\)}
=
\parbox{0.30\columnwidth}{\centering
Subgroup-Specific Specificity$_g$}
-
\parbox{0.30\columnwidth}{\centering
Subgroup-Specific Sensitivity$_g$}
\]

where a positive \(\delta_g\) indicates the model is conservative and favors the NCI cases, while a negative value shows that the model is aggressive and favors CI cases within the same subgroup. 

Inter-group performance disparity (\(\Delta\)) was computed as the difference in performance metrics across subgroups to measure representational bias:
\[
\parbox{0.10\columnwidth}{\centering
\(\Delta_{sens}\)}
=
\parbox{0.30\columnwidth}{\centering
Subgroup-Specific Sensitivity$_{group A}$}
-
\parbox{0.30\columnwidth}{\centering
Subgroup-Specific Sensitivity$_{group B}$}
\]
\[
\parbox{0.10\columnwidth}{\centering
\(\Delta_{spec}\)}
=
\parbox{0.30\columnwidth}{\centering
Subgroup-Specific Specificity$_{group A}$}
-
\parbox{0.30\columnwidth}{\centering
Subgroup-Specific Specificity$_{group B}$}
\]

Reporting both \(\delta\) and \(\Delta\) ensures that the analysis reflects both the nature of errors within each subgroup and equity of performance across them. 
Statistical significance of these disparities was assessed using paired t-tests (\textit{p\(<\)0.05}) across the five fixed cross-validation folds.

Furthermore, to isolate the source of performance disparities, we visualized model's score distributions using overlaid density histograms for each subgroup, where the overlap area between classes represents the region of maximum model uncertainty. 
The algorithmic bias is evidenced by a physical increase in class overlap, indicating the extracted features are less representative for that specific subgroup.
To quantify this separation power independent of the decision threshold, we computed the subgroup-specific Area Under the ROC Curve (AUC), which measures the probability that a randomly selected CI sample is ranked higher than a randomly selected NCI sample within a demographic:

\[AUC_g = P(Score(CI_g) > Score(NCI_g))\]

A significant gap in AUC across groups confirms that the disparity is not merely a calibration or thresholding artifact, but a fundamental difference in model's ability to represent and classify the data for those demographics. 

% --------------------------------------------------------------------------------------
\begin{table*}[t]
\centering
\caption{Top classification results for CI vs. NCI across acoustic feature types, dataset configurations, and classifiers. Results are reported as accuracy, UAR, sensitivity, and specificity, averaged over 5 fixed random data splits, and expressed as mean (standard deviation)\%. MFCCs and eGeMAPS are used as baseline features. The highest UAR for each dataset configuration is highlighted. Asterisks (*) indicate comparisons where paired t-tests yielded \textit{p\(<\)0.05}. Clf: classifier, IMB: imbalanced, CI-B: cognitive impairment balanced, CIG-B: cognitive impairment and gender balanced, Train-Bal/Test-Rem: train on balanced and test on remaining data, W2V-HL: Wav2Vec 2.0 hidden layer.}
\label{tab:ci-nci}
\resizebox{0.95\linewidth}{!}{
% \begin{tabular}{@{}ccccccc@{}}
\begin{tabular}{@{}>{\centering\arraybackslash}p{1.8cm} >{\centering\arraybackslash}p{2.3cm} >{\centering\arraybackslash}p{1.6cm}|>{\centering\arraybackslash}p{2.3cm} >{\centering\arraybackslash}p{2.3cm} >{\centering\arraybackslash}p{2.3cm} >{\centering\arraybackslash}p{2.3cm}@{}}
\toprule
\textbf{Dataset} & \textbf{Feature} & \textbf{Clf} & \textbf{Accuracy} & \textbf{UAR} & \textbf{Sensitivity} & \textbf{Specificity} \\ \midrule
IMB & \multirow{3}{*}{MFCCs} & \multicolumn{1}{c|}{MLP} & 66.38 (2.81) & 63.39 (3.36) & 77.14 (4.42) & 49.63 (7.98) \\
CI-B &  & \multicolumn{1}{c|}{MLP} & 63.33 (4.74) & 63.33 (4.74) & 65.18 (5.54) & 61.48 (6.02) \\
CIG-B &  & \multicolumn{1}{c|}{MLP} & 66.81 (3.95) & 66.75 (4.02) & 64.35 (10.79) & 69.17 (7.73) \\ \midrule
IMB & \multirow{3}{*}{eGeMAPS} & \multicolumn{1}{c|}{SVM} & 59.42 (2.25) & 49.73 (0.96) & 94.29 (7.16) & 5.19 (5.54) \\
CI-B &  & \multicolumn{1}{c|}{SVM} & 50.74 (5.31) & 50.74 (5.31) & 57.04 (6.87) & 44.45 (13.25) \\ 
CIG-B &  & \multicolumn{1}{c|}{SVM} & 46.81 (6.31) & 47.25 (6.31) & 68.69 (19.90) & 25.83 (19.61) \\
\midrule
IMB & \multirow{3}{*}{W2V2 HL 9} & \multicolumn{1}{c|}{MLP} & 81.16 (3.55) & \textbf{80.56 (5.02)*} & 83.33 (3.98) & 77.78 (12.61) \\
CI-B &  & \multicolumn{1}{c|}{MLP} & 80.00 (4.29) & \textbf{80.00 (4.29)*} & 77.04 (7.18) & 82.96 (4.45) \\
CIG-B &  & \multicolumn{1}{c|}{SVM} & 76.60 (3.01) & 76.54 (2.96)* & 73.91 (4.76) & 79.17 (6.97) \\ \midrule
IMB & \multirow{3}{*}{W2V2 HL 10} & \multicolumn{1}{c|}{SVM} & 80.00 (3.35) & 79.47 (2.99)* & 81.90 (7.31) & 77.04 (7.18) \\
CI-B &  & \multicolumn{1}{c|}{SVM} & 76.29 (1.81) & 76.29 (1.81)* & 74.81 (7.18) & 77.78 (6.19) \\
CIG-B &  & \multicolumn{1}{c|}{MLP} & 76.17 (4.54) & 76.18 (4.62)* & 76.52 (8.95) & 75.83 (3.11) \\ \midrule
\multicolumn{7}{c}{\textbf{Train-Bal/Test-Rem}} \\ \midrule
CIG-B & W2V2  HL 7 & \multicolumn{1}{c|}{MLP} & 76.83 (4.81) & \textbf{78.53 (2.76)} & 74.29 (10.77) & 82.78 (11.44) \\ \bottomrule
\end{tabular}}%
\end{table*}

\section{Experiments and results}
Two different experiments were performed in this work: (1) \textit{evaluation of feature performance in CI vs. NCI and D-CI vs. ND-CI classifications}, and (2) \textit{bias analysis of the best performing feature-classifier pairs across demographic and clinical factors}. 

\subsection{Feature Performance}
In this experiment the classification performance of three acoustic representations was assessed.
MFCCs and eGeMAPS were evaluated as baseline acoustic representations on all three dataset configurations i.e., imbalanced and CI-balanced, CI-gender balanced datasets.
Although embeddings from both the latent and hidden layers (numbered 1 to 12) of the W2V2 model were examined, only the highest-performing hidden layers are reported, as latent-layer embeddings consistently yielded lower classification performance.

Accuracy, UAR, sensitivity, and specificity were used as evaluation metrics.
Sensitivity reflects the proportion of CI (or depressed) cases correctly identified, while specificity reflects the proportion of NCI (or non-depressed) cases correctly identified \cite{bonvino2025digital}. 
To assess the reliability of performance differences, paired-t-tests were conducted on UARs from five fixed, stratified cross-validation folds.
Differences were considered statistically significant at a threshold of \textit{p\(<\)0.05}.

\subsubsection{CI vs. NCI}
% p-val paired t-test 0.0063 for Imb mfcc vs imb w2v9
% p-val paired t-test 0.00068 for imb egemaps vs imb w2vhl 9
The top classification results for each acoustic feature type, dataset configuration, and classifier are presented in Table \ref{tab:ci-nci}.
Among baseline features, MFCCs performed best, reaching a UAR of 63.3\% and 66.75\% on the CI-balanced and CI-gender balanced datasets with an MLP classifier.
Sensitivity and specificity were relatively balanced in both datasets: 65.2\% and 61.5\% in the CI-balanced dataset, and 64.3\% and 69.1\% in the CI-gender-balanced dataset.
On the imbalanced dataset, while UAR was comparable, sensitivity increased (77.1\%) at the cost of reduced specificity (49.6\%), indicating a tendency to over-classify CI.
This behavior is expected, as the imbalanced setting exposes the model to substantially more CI samples, leading it to prioritize CI detection while underrepresenting NCI patterns, whereas balancing restores equitable class representation and shifts the error distribution accordingly.
In comparison, eGeMAPS exhibited limited discriminative capacity, with its highest UAR reaching only 50.7\% on the CI-balanced dataset with an SVM classifier. 
This performance was notably poor on the imbalanced dataset, where classification results were skewed by extremely low specificity of 5.2\%, indicating substantial difficulty in correctly identifying NCI cases.

\begin{figure*}
    \centering
    \includegraphics[width=\linewidth]{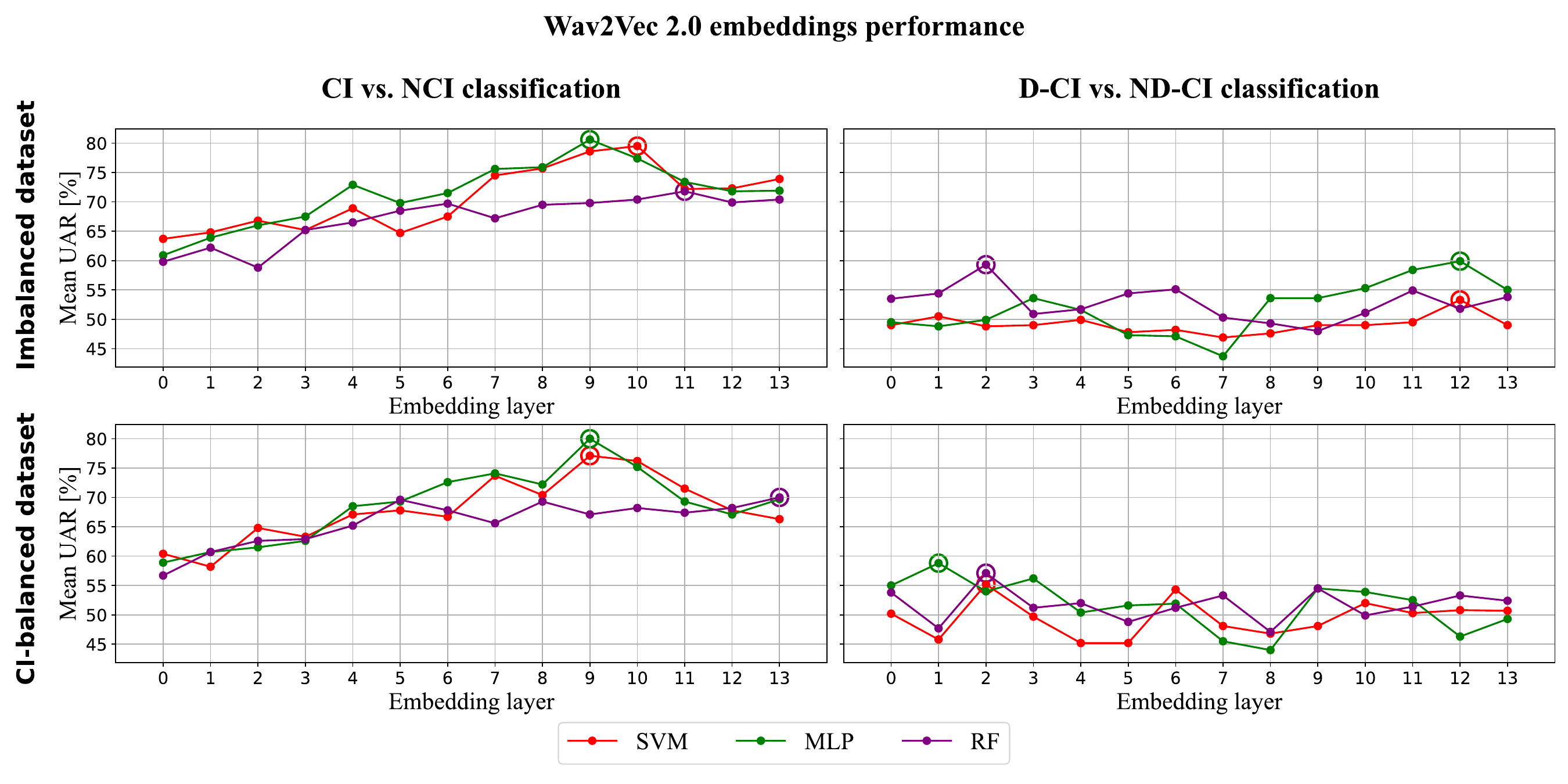}
    \caption{Comparison of classifier performance across W2V2 embedding layers in CI vs. NCI and D-CI vs. ND-CI classifications for imbalanced and CI-balanced datasets. Best results are highlighted for each classifier.}
    \label{fig:em_perf}
\end{figure*}

In contrast, W2V2 hidden layers 9 and 10 significantly outperformed traditional acoustic features across all dataset configurations (all \textit{p\(<\)0.05}).
W2V2 layer 9 combined with an MLP achieved the highest UAR of 80.6\% on the imbalanced dataset, with strong sensitivity (83.3\%) and specificity (77.8\%). 
Performance remained relatively consistent within each dataset configuration, with UAR=80.0\% on the CI-balanced dataset and UAR=76.5\% on the CI-gender balanced dataset.
Similarly, for W2V2 layer 10, imbalanced dataset yielded the highest accuracy of 80.0\% (UAR=79.5\%) using SVM, while maintaining consistent performance across other dataset configurations. 

Additionally, in an experimental setup where models were trained on the CI and gender-balanced dataset and evaluated on the remaining data, W2V2 layer 7 outperformed all other analyzed acoustic representations (UAR=78.5\%) with MLP, indicating strong generalization capability beyond balanced training conditions. 

Figure~\ref{fig:em_perf} demonstrates the comparative CI vs. NCI classification performance across W2V2 embedding layers for imbalanced and CI-balanced datasets.
Embedding layers 9 and 10 outperformed all the other layers for both datasets.
Furthermore, all classifiers demonstrated comparable performance up to embedding layer 6, beyond which SVM and MLP showed improvements in UARs but not the RF classifier.

\subsubsection{D-CI vs. ND-CI}
\begin{table*}[t]
\centering
\caption{Top classification results for D-CI vs. ND-CI across acoustic feature types, dataset configurations, and classifiers. Results are reported in terms of accuracy, UAR, sensitivity, and specificity, averaged over 5 fixed random data splits and expressed as mean (standard deviation)\%. MFCCs and eGeMAPS are used as baseline features. The highest UAR for each dataset configuration is highlighted. Clf: classifier, IMB: imbalanced, CI-B: cognitive impairment balanced, CIG-B: cognitive impairment and gender balanced, Train-Bal/Test-Rem: train on balanced and test on remaining data, W2V-HL: Wav2Vec 2.0 hidden layer.}
\label{tab:dci-ndci}
\resizebox{0.95\linewidth}{!}{
% \begin{tabular}{@{}ccccccc@{}}
\begin{tabular}{@{}>{\centering\arraybackslash}p{1.8cm} >{\centering\arraybackslash}p{2.3cm} >{\centering\arraybackslash}p{1.6cm}|>{\centering\arraybackslash}p{2.3cm} >{\centering\arraybackslash}p{2.3cm} >{\centering\arraybackslash}p{2.3cm} >{\centering\arraybackslash}p{2.3cm}@{}}
\toprule
\textbf{Dataset} & \textbf{Feature} & \textbf{Clf} & \textbf{Accuracy} & \textbf{UAR} & \textbf{Sensitivity} & \textbf{Specificity} \\ \midrule
IMB & \multirow{3}{*}{MFCCs} & \multicolumn{1}{c|}{RF} & 58.57 (5.13) & 51.78 (6.51) & 28.00 (12.22) & 75.56 (3.78) \\
CI-B &  & \multicolumn{1}{c|}{SVM} & 62.22 (1.48) & 50.65 (1.29) & 6.00 (12.00) & 95.29 (9.41) \\
CIG-B &  & \multicolumn{1}{c|}{MLP} & 58.33 (6.97) & 52.89 (6.76) & 31.11 (8.31) & 74.67 (8.84) \\ \midrule
IMB & \multirow{3}{*}{eGeMAPS} & \multicolumn{1}{c|}{MLP} & 55.24 (4.10) & 51.26 (3.74) & 37.33 (11.62) & 65.19 (9.83) \\
CI-B &  & \multicolumn{1}{c|}{SVM} & 61.48 (3.78) & 53.76 (5.15) & 24.00 (20.59) & 83.53 (13.62) \\
CIG-B &  & \multicolumn{1}{c|}{MLP} & 61.67 (9.65) & 59.11 (11.36) & 48.89 (19.37) & 69.33 (6.79) \\ \midrule
IMB & W2V2 HL 11 & \multicolumn{1}{c|}{MLP} & 63.33 (3.87) & 58.44 (3.34) & 41.33 (8.84) & 75.56 (8.31) \\ \midrule
IMB & W2V2 HL 12 & \multicolumn{1}{c|}{MLP} & 62.86 (8.73) & \textbf{59.85 (8.86)} & 49.33 (10.83) & 70.37 (9.37) \\ \midrule
CI-B & \multirow{2}{*}{W2V2 HL 2} & \multicolumn{1}{c|}{RF} & 61.48 (6.46) & 57.06 (7.57) & 40.00 (12.65) & 74.12 (4.71) \\
CIG-B &  & \multicolumn{1}{c|}{RF} & 66.67 (7.45) & \textbf{61.33 (7.84)} & 40.00 (11.33) & 82.67 (8.00) \\ \midrule
CI-B & \multirow{2}{*}{W2V2 HL 6} & \multicolumn{1}{c|}{SVM} & 62.22 (2.77) & \textbf{54.35 (2.70)} & 24.00 (16.25) & 84.71 (12.67) \\
CIG-B &  & \multicolumn{1}{c|}{RF} & 60.00 (4.25) & 50.22 (3.40) & 11.11 (0.00) & 89.33 (6.80) \\ \midrule
\multicolumn{7}{c}{\textbf{Train-Bal/Test-Rem}} \\ \midrule
CIG-B & W2V2 HL 2 & \multicolumn{1}{c|}{MLP} & 64.47 (5.54) & 57.89 (2.88) & 35.55 (13.93) & 82.22 (15.65) \\
CIG-B & W2V2 HL 6 & \multicolumn{1}{c|}{MLP} & 65.41 (7.46) & 56.43 (4.85) & 23.23 (8.01) & 89.63 (15.30) \\ \bottomrule
\end{tabular}}
\end{table*}

The top results of depression classification across all acoustic features and dataset configuration are summarized in Table~\ref{tab:dci-ndci}. 
Both MFCCs and eGeMAPS, demonstrated weak classification performance with UARs around 50\% and sensitivity as low as 6\% on the CI-balanced dataset, indicating substantial difficulty in reliably detecting depression within the CI population.

W2V2 embeddings slightly improved performance over baseline features features. 
On the imbalanced dataset, layer 12 with MLP achieved the highest UAR of 59.9\%, approximately 9\% higher than MFCCs, with moderate sensitivity (49.3\%).
While this represents a relative improvement, overall classification performance remained suboptimal. 
On the CI-balanced and CI-gender balanced datasets, layer 2 with an RF classifier achieved the highest UARs of 57.1\% and 61.3\%, respectively. 
However, across all configurations, higher specificity consistently came at the cost of reduced sensitivity, reflecting a tendency to under-identify depression while minimizing false positives.

To assess generalization, models trained on the CI-gender balanced dataset and evaluated on the remaining data produced inconsistent results. 
W2V2 layer 2 combined with an MLP classifier showed a decline in UAR of approximately 3.4\%, whereas layer 6 improved by 6.2\%, suggesting differences in layer-wise robustness. 

The comparative classification performance across W2V2 embedding layers for D-CI vs. ND-CI classification on the imbalanced and CI-balanced datasets is illustrated in Figure~\ref{fig:em_perf}. 
Performance trends were largely consistent across classifiers.
Embedding layer 2 yielded slight improvements in UAR for the RF classifier in both imbalanced and CI-balanced settings.
Additionally, embedding layers 11 and 12 achieved relatively better UARs on the imbalanced dataset. 
One likely explanation is the disproportionate class distribution in the imbalanced dataset (51 D-CI and 88 ND-CI cf.~Table~\ref{tab:data_distribution}), where overall UAR is more influenced by classification performance on the majority ND-CI class.    

Further exploratory analyses involved training models to predict CI status and testing them on depression labels, and vice versa, using the CI and gender-balanced dataset. 
These cross-condition evaluations yielded unsatisfactory results, with performance near chance levels (see Appendix Table~\ref{tab:trainx_testy}), highlighting minimal representational overlap between the two conditions.

% In summary, higher-order W2V2 embeddings consistently achieved superior performance in CI vs. NCI classification, while lower and mid-level layers yielded better results for D-CI vs. ND-CI classification.

\subsection{Bias Analysis}
\begin{table*}[t]
\centering
\caption{Subgroup performance and disparity metrics for CI vs.\ NCI classification using Wav2Vec 2.0 hidden layer 9 with an SVM classifier. $\delta$ denotes Intra-group performance imbalance (Sp - Se); positive values reflect bias toward NCI (higher specificity), while negative values indicate bias toward CI (higher sensitivity). $\Delta$ quantifies Inter-group performance disparity.Asterisks (*) indicate comparisons where paired t-tests yielded \textit{p\(<\)0.05}. All values are reported as percentages, with biased results highlighted. Se: Subgroup-specific sensitivity, Sp:Subgroup-specific specificity, IMB: imbalanced, CI-B: cognitive impairment balanced, CIG-B: cognitive impairment and gender balanced, N-Depr.: non-depressed, Depr.: depressed.}
\label{tab:bias-hl9}
\setlength{\tabcolsep}{4pt} % Maximum compression for horizontal fit
\begin{tabular}{@{}c | ccc | ccc | cc | ccc | ccc | cc | ccc | ccc | cc@{}}
\toprule
\multirow{3}{*}{\textbf{Dataset}} & \multicolumn{8}{c|}{\textbf{Age Groups}} & \multicolumn{8}{c|}{\textbf{Gender}} & \multicolumn{8}{c}{\textbf{Clinical Status}} \\ \cmidrule(lr){2-9} \cmidrule(lr){10-17} \cmidrule(l){18-25}
 & \multicolumn{3}{c|}{\textbf{Group 1}} & \multicolumn{3}{c|}{\textbf{Group 2}} & \multicolumn{2}{c|}{\textbf{Age $\Delta$}} & \multicolumn{3}{c|}{\textbf{Male}} & \multicolumn{3}{c|}{\textbf{Female}} & \multicolumn{2}{c|}{\textbf{Gender $\Delta$}} & \multicolumn{3}{c|}{\textbf{N-Depr.}} & \multicolumn{3}{c|}{\textbf{Depr.}} & \multicolumn{2}{c}{\textbf{Clin. $\Delta$}} \\
 & \textbf{Sp} & \textbf{Se} & \textbf{$\delta$} & \textbf{Sp} & \textbf{Se} & \textbf{$\delta$} & \textbf{$\Delta_{Sp}$} & \textbf{$\Delta_{Se}$} & \textbf{Sp} & \textbf{Se} & \textbf{$\delta$} & \textbf{Sp} & \textbf{Se} & \textbf{$\delta$} & \textbf{$\Delta_{Sp}$} & \textbf{$\Delta_{Se}$} & \textbf{Sp} & \textbf{Se} & \textbf{$\delta$} & \textbf{Sp} & \textbf{Se} & \textbf{$\delta$} & \textbf{$\Delta_{Sp}$} & \textbf{$\Delta_{Se}$} \\ \midrule
IMB & 80 & 76 & 4 & 71 & 83 & \textbf{-12} & \textbf{9} & -7 & 86 & 76 & \textbf{10$^{\ast}$} & 68 & 83 & \textbf{-15$^{\ast}$} & \textbf{18$^{\ast}$} & \textbf{-7} & 73 & 81 & -8 & 95 & 81 & \textbf{14$^{\ast}$} & \textbf{-22$^{\ast}$} & 0 \\
CIB & 84 & 78 & 6 & 69 & 75 & \textbf{-6} & \textbf{15$^{\ast}$} & 3 & 85 & 69 & \textbf{16$^{\ast}$} & 72 & 80 & \textbf{-8$^{\ast}$} & \textbf{13$^{\ast}$} & \textbf{-11$^{\ast}$} & 74 & 77 & -3 & 100 & 74 & \textbf{26$^{\ast}$} & \textbf{-26$^{\ast}$} & 3 \\
CIG-B & 79 & 74 & 5 & 80 & 74 & 6 & -1 & 0 & 86 & 74 & \textbf{12$^{\ast}$} & 73 & 74 & -1 & \textbf{13$^{\ast}$} & 0 & 76 & 77 & -1 & 100 & 70 & \textbf{30$^{\ast}$} & \textbf{-24$^{\ast}$} & 7 \\ \bottomrule
\end{tabular}
\end{table*}

\begin{figure*}
    \centering
    \includegraphics[width=0.9\linewidth]{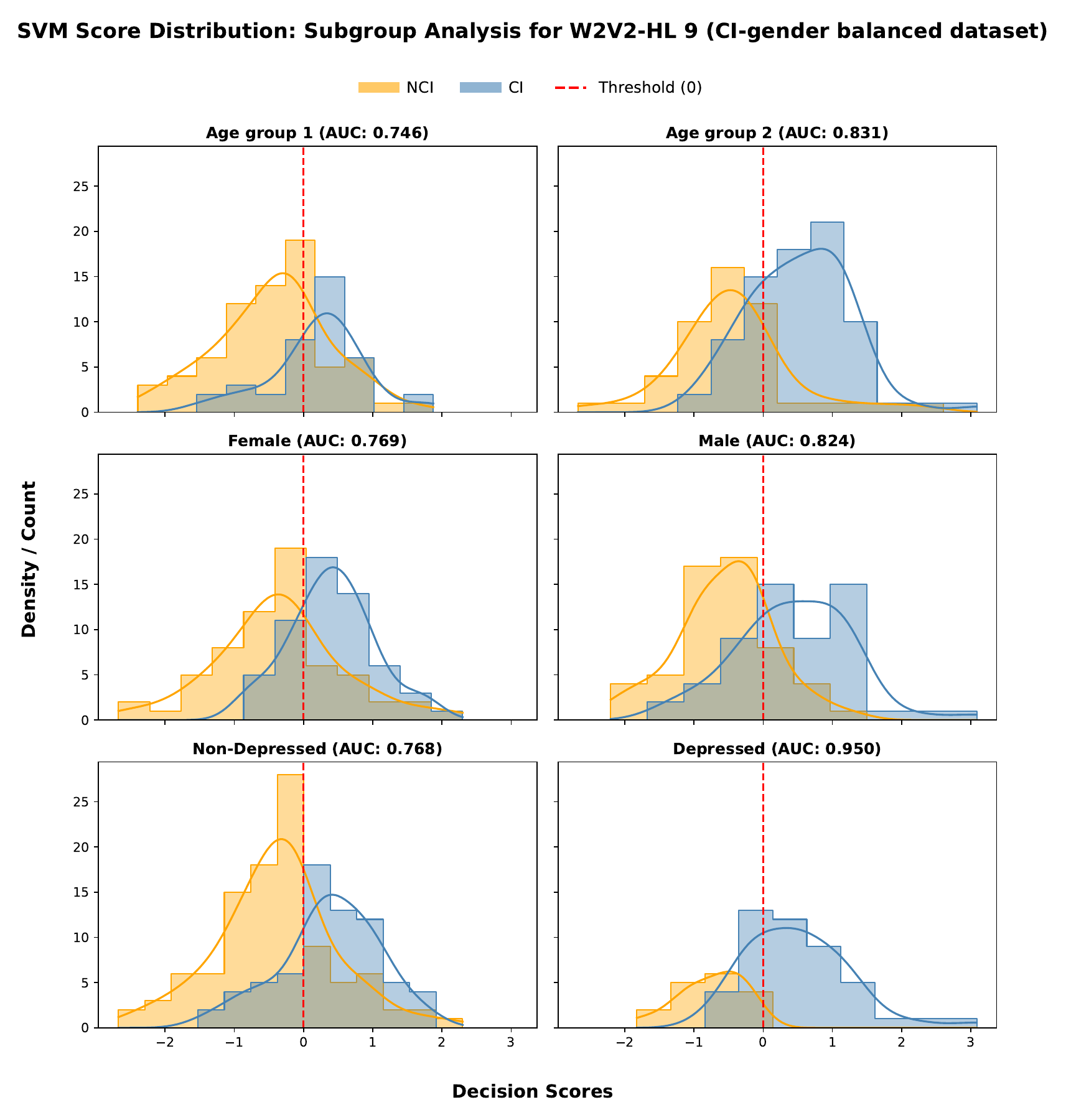}
    \caption{SVM decision score distributions across demographic and clinical subgroups for Wav2Vec 2.0 hidden layer 9 using CI-gender balanced dataset. Histograms and Kernel Density Estimate (KDE) curves illustrate the separation between NCI and CI classes. The dashed line at 0 represents the SVM decision threshold. AUC values are provided for each subgroup to indicate classification performance.}
    \label{fig:bias-hist}
\end{figure*}
To assess subgroup-level performance disparities in CI vs. NCI classification, the best-performing configurations across all datasets were utilized. 
Table~\ref{tab:bias-hl9} presents subgroup performance and disparity metrics for CI vs. NCI classification using W2V2 layer 9 with an SVM classifier across each demographic and clinical subgroup.
To further isolate the source of these performance disparities, model's score distributions are also analyzed as displayed in Figure~\ref{fig:bias-hist}.

The Intra-group performance imbalance \(\delta\) (Subgroup-specific Specificity - Subgroup-specific Sensitivity) captures trade-off between sensitivity and specificity within a subgroup, and Inter-group performance disparity \(\Delta\) measures differences in model performance across subgroups for a given metric (sensitivity or specificity). 
High \(\Delta\) and large AUC gaps is the primary indicator of algorithmic bias, indicating that the underlying acoustic features are less discriminative for the specific subgroup.
Whereas, high \(\Delta\) but similar AUC suggests a thresholding or calibration issue that can be mitigated via post-hoc scaling.

A systematic gender-based performance disparity was observed across all dataset configurations.
Intra-group analysis revealed a significant trade-off imbalance (\(\delta\)): male participants were consistently classified with higher specificity (favoring NCI detection), with \(\delta\) values of +10\% to +16\% (\textit{p\(<\)0.05}). 
Conversely, female participants exhibited a pronounced tilt toward sensitivity (favoring CI detection), particularly in the imbalanced and CI-balanced datasets (\(\delta \) = -15\% and -8\%, respectively; both \textit{p\(<\)0.05}).

This internal imbalance also translated into substantial inter-group performance disparities (\(\Delta\)).
The model demonstrated a systematic advantage for males in specificity (\(\Delta_{spec}\) ranging from +13\% to +18\%, \textit{p\(<\)0.05}), indicating that healthy females are significantly more likely to be misclassified than their males counterparts. 
This is also supported by the subgroup-specific AUCs, where males achieved higher discriminative power (\(AUC_{Male}\) = 0.824) compared to females (\(AUC_{Female}\) = 0.769).
The higher histogram overlap in the female subgroup suggests an inherent algorithmic bias, where model struggles to distinguish classes in female speech regardless of the decision threshold.

Analysis across age demographics highlighted a clear performance gap favoring older participants (age group 2).
While the model maintained a relatively balanced intra-group performance for age group 1 (\(\delta\) ranging from +4\% to +6\%), age group 2 exhibited a strong aggressive bias toward CI detection (\(\delta\) as low as -12\%).

Inter-group investigation revealed high specificity gap (\(\Delta_{spec}\) as high as +15\%, \textit{p\(<\)0.05}) for age group 1 favoring NCI detection.
This disparity is fundamentally driven by a difference in model's discriminative capacity (algorithmic bias); age group 1 \(AUC\) was limited to 0.746 as compared to age group 2 \(AUC\) of 0.831.
The density plots indicate a larger class overlap for the younger (age group 1) subgroup, suggesting that the underlying acoustic features are less representative of CI-related changes in younger participants.

The strongest and most concerning performance disparity was observed in participants with depression.
The intra-group analysis revealed balanced performance of non-depressed participants (\(\delta\) ranging from -1\% to -8\%).
Whereas depressed participants exhibited a conservative bias toward NCI detection with \(\delta\) as high as +30\%, \textit{p\(<\)0.05}), even after data balancing.

The model exhibited high discriminative performance for depressed individuals (\(AUC_{Depressed}\) = 0.950), compared to a much lower \(AUC_{Non-depressed}\) = 0.768.
This algorithmic advantage translated to a high specificity gap (\(\Delta_{spec}\) as low as -26\%, \textit{p\(<\)0.05}).
The score distributions for the depressed participants showed low overlap, whereas the non-depressed subgroup exhibited high density overlap.

Moreover, a comparative bias analysis of W2V2 hidden layer 10 with an MLP classifier was performed (cf.~Appendix~Table~\ref{tab:bias-hl10}).
Similar patterns of subgroup performance disparities and bias were found particularly for females and depressed participants.
Importantly, no bias analysis was conducted for the D-CI vs. ND-CI classification task, as classification performance was unsatisfactory (UARs around 50\%), precluding meaningful subgroup comparisons.

In summary, higher-layer W2V2 embeddings substantially outperform traditional acoustic features for CI detection, but exhibit clear demographic and clinical biases. 
Depression classification remains a more challenging task, with limited generalization between CI and depression, underscoring the need for condition-specific representations.

% --------------------------------------------------------------------------------------
\section{Discussion}
This study evaluated three acoustic representations, including MFCCs, eGeMAPS, and W2V2, with multiple classifiers for detecting CI and depression among CI participants.

For CI classification, MFCCs demonstrated moderate performance, while eGeMAPS exhibited notably poor specificity, dropping to 5.2\%, highlighting its limited effectiveness for CI detection. 
In contrast, W2V2 embeddings from higher-order layers 9 and 10 consistently achieved superior performance, with UARs between 76.5\% and 80.6\% across all dataset configurations and classifiers. 
These improvements likely stem from the ability of deeper transformer layers to encode contextual and semantic speech features~\cite{pasad2021layer}, which are essential for detecting linguistic and cognitive disruptions characteristic of CI~\cite{fraser2015linguistic}. 
Their stable performance across CI- and gender-balanced datasets underscores their robustness.
Several studies have reported accuracies of up to 82\%, which aligns well with our results demonstrating accuracy of up to 81.56\%~\cite{paula, braun, balagopalan}.

The depression classification task proved more challenging.
Both MFCCs and eGeMAPS showed unsatisfactory performance, with UARs around 50\%.
While W2V2 embeddings slightly improved results, overall performance remained suboptimal, with the highest UAR reaching only 61.3\% (layer 2 with RF classifier).
Among W2V2 embeddings, lower and mid-level layers (particularly layers 2 and 6) outperformed higher layers, which aligns with findings that acoustic markers of depression, such as monotone pitch, reduced prosodic variation, and increased jitter and shimmer are encoded in earlier layers~\cite{low2020automated, pasad2021layer}.
Given these consistently low performances, we refrain from drawing strong conclusions from depression classification and did not extend bias assessments to this task, as such evaluations would not be meaningful without a sufficiently reliable baseline model.

Attempts to model the overlap between CI and depression yielded poor generalization. 
Classifiers trained to detect CI failed to transfer to depression detection, and vice versa, with performance near chance levels (cf. Appendix Table~\ref{tab:trainx_testy}).
This suggests that while CI and depression may share clinical symptoms, their acoustic manifestations are distinct, necessitating separate modeling strategies.

Bias analysis using W2V2 layer 9 with an SVM classifier revealed consistent disparities across all datasets.
Intra-group imbalance (\(\delta\)) and inter-group disparity (\(\Delta\)), alongside subgroup-specific AUCs were utilized, to isolate algorithmic bias from simple thresholding effects.
A systematic gender bias was observed, with males consistently classified with higher specificity (\(\delta\) up to +16\%, favoring NCI detection) while females exhibited a tilt toward sensitivity (\(\delta\) as low as -15\%, favoring CI detection). 
This resulted in substantial inter-group specificity disparities (\(\Delta_{spec}\) up to +18\%, \textit{p\(<\)0.05}). 
Critically, the lower discriminative power for females (\(AUC_{Female}\) = 0.769 vs. \(AUC_{Male}\) = 0.824) and higher histogram overlap (Figure \ref{fig:bias-hist}) point to an inherent algorithmic bias. 
This suggests that W2V2 embeddings may struggle to distinguish CI-related features in female speech, a trait likely inherited from imbalanced pretraining data like LibriSpeech \cite{baevski20}.

Age-related analysis highlighted a clear performance gap favoring older adults (age group 2).
While age group 2 exhibited an aggressive bias toward CI detection (\(\delta\) = -12\%), group 1 was limited by a lower \(AUC\) (0.746 vs. 0.831). 
The larger class overlap in density plots for age group 1 confirms that the underlying acoustic features are less representative of CI-related changes in younger participants, leading to higher specificity gaps (\(\Delta_{spec}\) up to +15\%)~\cite{fraser2015linguistic}.
These results are in line with prior studies reporting gender and age-related disparities in speech SSL models~\cite{boito2022study, fuckner2023uncovering, lin2024social}.

The strongest disparity was observed regarding depression status. 
While non-depressed participants showed balanced performance, depressed individuals exhibited a sharp conservative bias toward NCI detection (\(\delta\) up to +30\%). 
However, the near-perfect separation in score distributions and superior discriminative power for depressed individuals (\(AUC_{Depressed}\) = 0.950 vs. \(AUC_{Non-depressed}\) = 0.768) reveal a significant clinical status bias, likely due to the low number of depressed NCI individuals in the dataset (cf.~Table~\ref{tab:data_distribution}).
Consistent trends across W2V2 layer 10 with an MLP classifier suggest these representational biases are not confined to a single layer or classifier.
However, in the CI-gender balanced dataset, inconsistent results for depressed NCI subjects, compared to layer 9, likely reflect the small sample size, where one misclassification can substantially impact accuracy (Table~\ref{tab:bias-hl10}).  

These findings highlight a key trade-off in clinical speech applications: while balancing datasets can mitigate some disparities, it may not fully eliminate biases embedded in model representations or data. 
This supports the argument for fairness-aware model development and evaluation rather than solely relying on data balancing to ensure equitable clinical decision support. 
Importantly, our work extends previous studies by explicitly evaluating bias in acoustic-based ML models for CI detection across demographic and clinical subgroups, a gap previously highlighted in the literature~\cite{lin2024social}. 

However, this study focused exclusively on the DB's Pitt Corpus, which may not fully represent the demographic and linguistic diversity of broader clinical populations.
This limitation could affect the generalizability of our findings. 
Additionally, a key limitation of this study lies in the limited number of depression-labeled samples available in the Pitt corpus \cite{becker1994natural}. 
As this corpus was not primarily designed for depression research, the limited representation of depressive cases may introduce variability in the results and constrain the robustness of conclusions related to depression-specific classification performance.

Future work should address these disparities by incorporating larger, more diverse, and better-balanced datasets with clinically validated depression annotations. 
It should also conduct subject-level analyses to more comprehensively characterize variability in clinical speech.
Extending this analysis to other modern SSL variants (e.g., HuBERT, WavLM, W2V-BERT-2.0) would further clarify how different pretraining paradigms influence bias.
Additionally, adaptive weighting and fusion of multiple embedding layers represents a promising direction for future work, particularly once the underlying layer-wise bias patterns are better understood.
Finally, it should explore bias-reduction methods to mitigate subgroup differences and support equitable, reliable model evaluation and the responsible deployment of AI in healthcare.
% --------------------------------------------------------------------------------------
\section{Conclusion}
This study examined the effectiveness of acoustic representations for classifying CI and comorbid depression, with a particular focus on identifying performance disparities across demographic and clinical subgroups.
The findings demonstrated that W2V2 embeddings, especially from higher transformer layers, consistently outperformed baseline features (MFCCs and eGeMAPS) for CI detection, achieving UARs of up to 80.6\%.
In contrast, depression classification within CI participants was more challenging, with lower and mid-level W2V2 layers yielding relatively better results. 
Cross-condition analyses revealed poor generalization between CI and depression, indicating that despite some overlapping symptoms, their acoustic patterns differ substantially and thus require task-specific modeling strategies

Bias analysis revealed performance disparities across demographic and clinical subgroups.
Specifically, within females and younger participants, the model exhibited pronounced class overlap and reduced discriminative power (\(AUC_{Female}\) = 0.769; \(AUC_{Age group 1}\) = 0.746), leading to substantial inter-group specificity disparities (\(\Delta_{spec}\) up to +18\%).
In contrast, male and older participants and participants with depression reached significantly higher discriminative accuracy with less class overlap.
These disparities highlight critical limitations in current speech-based CI detection models, indicating that even state-of-the-art approaches may inadvertently reflect and perpetuate biases embedded in training data. 

While our analysis focuses on a single dataset, the depth of subgroup-level evaluation provides insights into how bias may arise in self-supervised models. 
Future work should extend this analysis to larger and more diverse datasets with better-balanced clinically validated depression samples, enabling meaningful subject-level investigations to better capture variability in clinical speech. 
Additionally, it should systematically evaluate model debiasing strategies, to improve fairness and robustness, and explore the bias characteristics of other SSL architectures such as HuBERT, WavLM and W2V-BERT-2.0.
Overall, this study demonstrates that while self-supervised models like W2V2 can advance clinical speech analysis, they may also encode biases that reflect systemic inequities in training data. 
Speech-based diagnostic tools must therefore be evaluated not only for accuracy, but for fairness, especially when deployed in diverse clinical populations.

\appendices
\section*{Appendix}
\begin{table*}[t]
\centering
\caption{Classification performance for cross-condition evaluations on the CI and gender-balanced dataset. Models trained on CI status were tested on depression status and vice versa. Results are reported in terms of accuracy, UAR, sensitivity, and specificity, averaged over 5 fixed random data splits and expressed as mean (standard deviation)\%. Clf: classifier, Train-CI Test-D: train on CI and test for depression status, Train-D Test-CI: train on depression and test for CI status, W2V-HL: Wav2Vec 2.0 hidden layer.}
\label{tab:trainx_testy}
% \begin{tabular}{@{}ccc|cccc@{}}
\resizebox{0.9\linewidth}{!}{
% \begin{tabular}{@{}ccccccc@{}}
\begin{tabular}{@{}>{\centering\arraybackslash}p{1.8cm} >{\centering\arraybackslash}p{2cm} >{\centering\arraybackslash}p{1.3cm}|>{\centering\arraybackslash}p{2cm} >{\centering\arraybackslash}p{2cm} >{\centering\arraybackslash}p{2cm} >{\centering\arraybackslash}p{2cm}@{}}
\toprule
\textbf{Task} & \textbf{Feature} & \textbf{Clf} & \textbf{Accuracy} & \textbf{UAR} & \textbf{Sensitivity} & \textbf{Specificity} \\ \midrule
\multirow{3}{*}{\begin{tabular}[c]{@{}c@{}}Train-CI\\ Test-D\end{tabular}} & \begin{tabular}[c]{@{}c@{}}W2V\\ HL 7\end{tabular} & SVM & 48.70 (7.48) & \textbf{52.37 (6.13)} & 69.56 (6.81) & 35.18 (9.46) \\
 & \begin{tabular}[c]{@{}c@{}}W2V\\ HL 9\end{tabular} & RF & 45.22 (6.51) & 49.31 (6.03) & 67.69 (14.10) & 30.93 (16.01) \\
 & \begin{tabular}[c]{@{}c@{}}W2V\\ HL 10\end{tabular} & RF & 46.09 (6.51) & 49.64 (4.75) & 67.46 (9.79) & 31.81 (13.93) \\ \midrule
\multirow{2}{*}{\begin{tabular}[c]{@{}c@{}}Train-D\\ Test-CI\end{tabular}} & \begin{tabular}[c]{@{}c@{}}W2V\\ HL 2\end{tabular} & MLP & 52.34 (5.65) & 51.76 (5.61) & 38.00 (9.54) & 65.53 (10.84) \\
 & \begin{tabular}[c]{@{}c@{}}W2V\\ HL 6\end{tabular} & MLP & 57.45 (2.33) & \textbf{55.86 (3.09)} & 29.41 (5.32) & 82.30 (3.75) \\ \bottomrule
\end{tabular}}
\end{table*}

\begin{table*}[t]
\centering
\caption{Subgroup performance and disparity metrics for CI vs.\ NCI classification using Wav2Vec 2.0 hidden layer 10 with an MLP classifier. $\delta$ denotes Intra-group performance imbalance (Sp - Se); positive values reflect bias toward NCI (higher specificity), while negative values indicate bias toward CI (higher sensitivity). $\Delta$ quantifies Inter-group performance disparity.Asterisks (*) indicate comparisons where paired t-tests yielded \textit{p\(<\)0.05}. All values are reported as percentages, with biased results highlighted. Se: Subgroup-specific sensitivity, Sp:Subgroup-specific specificity, IMB: imbalanced, CI-B: cognitive impairment balanced, CIG-B: cognitive impairment and gender balanced, N-Depr.: non-depressed, Depr.: depressed.}
\label{tab:bias-hl10}
\setlength{\tabcolsep}{4pt}
\begin{tabular}{@{}c | ccc | ccc | cc | ccc | ccc | cc | ccc | ccc | cc@{}}
\toprule
\multirow{3}{*}{\textbf{Dataset}} & \multicolumn{8}{c|}{\textbf{Age Groups}} & \multicolumn{8}{c|}{\textbf{Gender}} & \multicolumn{8}{c}{\textbf{Clinical Status}} \\ \cmidrule(lr){2-9} \cmidrule(lr){10-17} \cmidrule(l){18-25}
 & \multicolumn{3}{c|}{\textbf{Group 1}} & \multicolumn{3}{c|}{\textbf{Group 2}} & \multicolumn{2}{c|}{\textbf{Age $\Delta$}} & \multicolumn{3}{c|}{\textbf{Male}} & \multicolumn{3}{c|}{\textbf{Female}} & \multicolumn{2}{c|}{\textbf{Gender $\Delta$}} & \multicolumn{3}{c|}{\textbf{N-Depr.}} & \multicolumn{3}{c|}{\textbf{Depr.}} & \multicolumn{2}{c}{\textbf{Clin. $\Delta$}} \\
 & \textbf{Sp} & \textbf{Se} & \textbf{$\delta$} & \textbf{Sp} & \textbf{Se} & \textbf{$\delta$} & \textbf{$\Delta_{Sp}$} & \textbf{$\Delta_{Se}$} & \textbf{Sp} & \textbf{Se} & \textbf{$\delta$} & \textbf{Sp} & \textbf{Se} & \textbf{$\delta$} & \textbf{$\Delta_{Sp}$} & \textbf{$\Delta_{Se}$} & \textbf{Sp} & \textbf{Se} & \textbf{$\delta$} & \textbf{Sp} & \textbf{Se} & \textbf{$\delta$} & \textbf{$\Delta_{Sp}$} & \textbf{$\Delta_{Se}$} \\ \midrule
IMB & 79 & 78 & 1 & 73 & 79 & -6 & 6 & -1 & 75 & 76 & -1 & 78 & 80 & \textbf{-2} & -3 & -4 & 73 & 77 & -4 & 95 & 81 & \textbf{14$^{\ast}$} & \textbf{-22$^{\ast}$} & -4 \\
CIB & 80 & 78 & 2 & 75 & 70 & \textbf{5} & 5 & 8 & 78 & 69 & \textbf{9$^{\ast}$} & 78 & 75 & 3 & 0 & -6 & 74 & 75 & -1 & 100 & 67 & \textbf{33$^{\ast}$} & \textbf{-26$^{\ast}$} & 8 \\
CIG-B & 72 & 74 & -2 & 82 & 78 & \textbf{4} & -10 & -4 & 82 & 74 & \textbf{8$^{\ast}$} & 70 & 79 & \textbf{-9$^{\ast}$} & \textbf{12$^{\ast}$} & -5 & 74 & 77 & -3 & 88 & 76 & \textbf{12$^{\ast}$} & \textbf{-14$^{\ast}$} & 1 \\ \bottomrule
\end{tabular}
\end{table*}

This appendix provides additional evaluation results and subgroup analyses to supplement the findings reported in the main paper.

Table~\ref{tab:trainx_testy} summarizes the results of cross-condition evaluations where models trained to classify CI were tested on depression labels and vice versa. 
The results demonstrate poor generalization across conditions, with accuracies and UARs hovering around 50\% confirming that despite some symptomatic overlap, CI and depression manifest distinct acoustic patterns not effectively captured by shared models.

Table~\ref{tab:bias-hl10} reports subgroup-specific sensitivity and specificity for CI vs. NCI using W2V2 layer 10 with an MLP classifier across different dataset configurations. 
Similar to the analysis for layer 9, notable biases are observed, particularly a persistent bias toward NCI in depressed participants (\(\Delta_{spec}\) as low as -26\%) and variable performance disparities across gender and age groups, further underscoring representational and distributional imbalances in speech-based CI detection.
These findings further highlight the importance of systematic bias evaluation in clinical speech-based AI models to ensure equitable and reliable diagnostic support.

\bibliographystyle{IEEEtran}
\bibliography{literature}
\end{document}